\documentclass[journal]{IEEEtran}

\usepackage{units}
\usepackage{gensymb}
\usepackage{bm}
\usepackage{todonotes}
\usepackage{graphicx,epstopdf}
\usepackage{float}
\usepackage{amsmath}
\usepackage{enumitem}
\usepackage{xcolor}

\usepackage{soul} % To highlight text

\usepackage{multirow}
\usepackage{microtype} % For command \textls[]{}
\usepackage{tikz} % For \foreach used for Orcid icon
\usepackage{totcount} % To enable extracting the value of the counter "page" 
\usepackage{stfloats}
\usepackage{hyperref}
\usepackage{hhline}

\graphicspath{{img/}}

\begin{document}

\title{Novel IMU-based Adaptive Estimator of the Center of Rotation of Joints for Movement Analysis}
\author{Sara García-de-Villa, Ana Jiménez-Martín and J. Jesús García-Domínguez}% <-this % stops a space
\markboth{Journal of \LaTeX\ Class Files,~Vol.~14, No.~8, September~2020}%
{Shell \MakeLowercase{\textit{et al.}}: Bare Demo of IEEEtran.cls for IEEE Journals}
\maketitle

\begin{abstract}
The location of the center of rotation (COR) of joints is a key parameter in multiple applications of human motion analysis.
The aim of this work was to propose a novel real-time estimator of the center of fixed joints using an inertial measurement unit (IMU).  
Since the distance to this center commonly varies during the joint motion due to soft tissue artifacts (STA), our approach is aimed at adapting to these small variations when the COR is fixed.
Our proposal, called ArVE$\mathbf{_\textit{d}}$, to the best of our knowledge, is the first real-time estimator of the IMU-joint center vector based on one IMU.
Previous works are off-line and require a complete measurement batch to be solved and most of them are not tested on the real scenario. 
The algorithm is based on an Extended Kalman Filter (EKF) that provides an adaptive vector to STA motion variations at each time instant, without requiring a pre-processing stage to reduce the level of noise. 
ArVE$\mathbf{_\textit{d}}$ has been tested through different experiments, including synthetic and real data. 
The synthetic data are obtained from a simulated spherical pendulum whose COR is fixed, considering both a constant and a variable IMU-joint vector, that simulates translational IMU motions due to STA. 
The results prove that ArVE$\mathbf{_\textit{d}}$ is adapted to obtain a vector per sample with an accuracy of \unit[$\mathbf{6.8\pm3.9}$]{mm} on the synthetic data, that means an error lower than \unit[$\mathbf{3.5}$]{\%} of the simulated IMU-joint vector. 
Its accuracy is also tested on the real scenario estimating the COR of the hip of $\mathbf{5}$ volunteers using as reference the results from an optical system. 
In this case, ArVE$\mathbf{_\textit{d}}$ gets an average error of \unit[$\mathbf{9.5}$]{\%} of the real vector value. In all the experiments, ArVE$\mathbf{_\textit{d}}$ outperforms the published results of the reference algorithms.

\end{abstract}


\IEEEpeerreviewmaketitle

\section{Introduction}
\label{sec:intro}

Human motion analysis is a fundamental support tool to obtain objective information about the movement parameters, which are especially important in sports or clinical rehabilitation routines and preventive treatments~\cite{Fitzgerald2007}. 
With the current ageing in developed countries, the demand for home rehabilitation has increased and, with it, the need to obtain quantitative exercise data remotely.
Optical motion capture is considered the \emph{gold standard} technique in the motion analysis field, but the systems that use it are very expensive and require high computational cost. Recently, low-cost optical methods have been developed, based on the analysis through different software applications of recordings captured by video cameras or smart mobile  devices~\cite{Mills2015}. However, optical systems are limited to controlled environments and suffer from occlusions, so wearable systems, such as inertial systems, have emerged as alternatives for motion analysis~\cite{kavanagh2008, Weygers2020, deBaets20}.
Focusing on the inertial capture systems, we can find two main alternatives: using an inertial measurement unit (IMU) attached to each segment of the human body  to measure its orientation~\cite{Allseits2017,Lin2012,bonnet2015,joukov2017rhythmic,elgohary2015, weygers2020drift, teufl2018validity}; or using one IMU to measure the orientation of the attached and the adjacent segments~\cite{bonnet2015}.

Some inertial systems require the characterization of personalized multi-body kinematic models in order  to describe movements. 
In this case, as IMU measurements are given in the sensor frame,  the relationships between the sensor and anatomical frames are needed for the calculation of the motion of the joints and segments.
Thus, most of algorithms aimed at calibrating IMUs with respect to the human body focus only on identifying the orientation of the  axes of rotation of joints with respect to the IMU axes, as in~\cite{muller17, cutti2008, favre2009calib}. 
However, recent studies have shown improvements in the accuracy of segments orientation estimation by exploiting the equations of motion for the translational acceleration of rotating bodies fusing information from several IMUs~\cite{Lin2012,bonnet2015,joukov2017rhythmic,elgohary2015, weygers2020drift, teufl2018validity}. 
In order to use this approach, the location of the  center of rotation (COR) of joints with respect to IMUs is needed.
In inertial systems, the COR is determined through the oriented vector $\bm{r}$, defined from the IMU accelerometers to the COR of joints, see Fig.~\ref{fig:expl_eqacels}.
The estimation of the orientation of joints is highly sensitive to the accuracy in obtaining this vector $\bm{r}$.

IMU-based algorithms have been proposed to determine this vector~\cite{seel2014, Crabolu2016,Frick2018,Schauer2017, Seel2012,mcginnis2013inertial, Olsson2017}. 
These algorithms can be separated between those that get an average $\bm{r}$ and those that obtain an adaptive $\bm{r}$. In the former, it is assumed that changes in $\bm{r}$ caused by soft tissue artifacts (STA) are eliminated by using signals of several seconds duration~\cite{mcginnis2013inertial,Crabolu2016,Crabolu2017}. However, these approaches lead to errors on scenarios with STA~\cite{Olsson2017, Frick2018}. 
As an alternative, Frick and Rahmatalla~\cite{Frick2018} propose an adaptive gradient descent method to obtain an $\bm{r}$ at each time instant in fixed CORs. 
This proposal is tested with synthetic data but it has not been tested on the real scenario of human joints.

In this work, we propose a novel estimator of $\bm{r}$ in fixed joints, which is adapted to variations in the relative positions of IMUs caused by STA. Our proposal is called \emph{ArVE$_d$}, that stands for Adaptive $\bm{r}$ Vector Estimator. This algorithm is based on the method introduced in~\cite{sgv_memea2020}. To evaluate the performance of our proposal, we compare ArVE$_d$ with the approach described in~\cite{Crabolu2016}, used as a reference to estimate an average $\bm{r}$, hereinafter  \emph{MrVS}, that stands for Mean $\bm{r}$ Vector least-Squares-based estimator.

Therefore,  the main goal of this work is to demonstrate that ArVE$_d$ is a competitive option to estimate the location of CORs of fixed human joints with one IMU, assuming that the IMU undergoes STA during motions.  The objectives are as follows:
	\begin{itemize}[leftmargin=*,labelsep=5.8mm]
		\item Validate the proposed algorithm, ArVE$_d$, with synthetic data.
		\item Evaluate the effect of the adaptation to variations in $\bm{r}$ comparing ArVE$_d$ with MrVS, which does not consider the effect of STA.
		\item Study the accuracy of ArVE$_d$ with real data of the COR of hips.
	\end{itemize}

This document has five sections besides the introduction. 
Section~\ref{sec:Related_works} includes a revision of the state of the art of inertial methods to obtain the location of CORs.
Section~\ref{sec:proposed_system} details the proposed ArVE$_d$ method together with our implementation of MrVS.
Section~\ref{sec:sim_experiments} explains the experiments with synthetic data and the achieved results of ArVE$_d$, compared with the results of our adaptation of MrVS.
Section~\ref{sec:real_experiments} describes the experiments on the real scenario of calibration of the hip center in five volunteers, together with the discussion of the results from ArVE$_d$ and MrVS, versus the results obtained with an optical system.
Finally, section~\ref{sec:conclusions} summarizes the main conclusions of this work.

\section{Related works}\label{sec:Related_works}

There are different works focused on the location of the COR of human joints, since the determination of an internal point of the body, as this center, is not trivial.
The most accurate approaches to determine the position and location of IMUs with respect to anatomical CORs or joint axes orientation are based on X-ray or magnetic resonance image, but both approaches are high priced, invasive and, ultimately, impractical~\cite{Crabolu2017}.
Therefore, CORs in the motion analysis field are commonly determined through palpation of external anatomic landmarks by expert therapists or by the use of optical systems~\cite{bonnet2015,joukov2017rhythmic,elgohary2015}.
Optical systems use sphere-fitting approaches to find the radius that best fits a trajectory described by optical markers~\cite{huang2000definitions, ehrig2006survey, DeRosario2014}. Both the palpation and the optical systems require expert hands to place the markers and are limited to controlled environments. These methods to obtain $\bm{r}$ use external information other than IMU-derived data, such as the location of the optical system markers. Thus, it limits the use of the biomechanical model-based inertial motion analysis to environments where optical systems are available.

As aforementioned, there are different proposals of IMU-based algorithms to determine this vector in the literature. 
In~\cite{Seel2012}, a method for estimating the location of knees, modelled as hinge joints, is introduced and tested by adding different levels of signal-to-noise ratio. This algorithm is also tested on the human gait scenario~\cite{seel2014}, but no conclusions on the absolute error in the $3$D joint location estimates are given. For locating fixed CORs instead of axes, McGinnis and Perkins propose in~\cite{mcginnis2013inertial} an algorithm based on exploiting the relationship between linear acceleration and turn rate in rigid solids. 
The algorithm is based on solving the equation of accelerations of a rigid-solid body moving in the $3$D space around a fixed COR whose linear acceleration is equal to zero. 
They reported an error of \unit[$3.1$]{mm} in tests with a mechanical analogue of the hip joint performing a determined joint motion (with a specific trajectory, range and velocity). The dependence of this method to different types of motions, ranges of motion position of IMUs and joint velocity is assessed in~\cite{Crabolu2016}. This study reports a considerable impact of the angular velocity on the COR identification and non-critical relations with the type and ranges of motion. 
This algorithm is tested on the glen-humeral joint estimation scenario~\cite{Crabolu2017},  
reporting an accuracy of \unit[$21$]{mm} compared with magnetic resonance images. It was concluded that the location of fixed CORs is more accurate using the information of one IMU in the algorithm of~\cite{Crabolu2016} than using data from two devices due to the small amplitude of the signals recorded by the IMU placed on the fixed segment and the difficulty related to its tracking.
Olsson and Halvorsen tested the same proposal in the case of moving CORs in mechanical simulations, studying different methodologies of solving the acceleration equation~\cite{Olsson2017}. However, the accuracy of this approach on the real scenario of human joints is not reported.

The previous algorithms obtain a mean value of $\bm{r}$ for each test, averaging the STA. 
However, the effect of STA is studied in~\cite{Olsson2017}  
the least squares method used in previous studies (such as~\cite{mcginnis2013inertial,Crabolu2016,Crabolu2017}) shows no robustness to outliers that may occur as a consequence of STA. And as stated in~\cite{Frick2018}, the STA can introduce significant errors in the location of the center of joints when assuming an average value of $\bm{r}$. 

To overcome this limitation associated with the STA, Frick and Rahmatalla propose a gradient descent method to obtain an $\bm{r}$ at each time instant with an IMU attached to a hinge joint~\cite{Frick2018}. However, this adaptive method requires its initialization using the complete test data with a duration around \unit[$25$]{s}, otherwise it may reach a local minimum.  
This proposal is tested with synthetic data from a $2$D-pendulum simulating STA with an attached spring and reports errors of \unit[$7.53$]{mm}.
In~\cite{Frick2018a}, the algorithm is evaluated with a mechanical hinge joint in which the effect of STA is replicated with the IMU placed on a piece of raw meat. The authors provide results on synthetic data, where the errors range from \unit[$10.8$]{mm} to \unit[$21.4$]{mm} on the highest STA scenarios. However, this algorithm has not been tested on the real scenario of human joints.

It is remarkable that all aforementioned inertial approaches entail different signal pre-processing to reduce noise in the IMU data and to estimate the angular acceleration, $\dot{\bm{\omega}}$, a common parameter required in these methods. 
In contrast, our initial proposed method, called ArVE, estimates an $\bm{r}$ at each time instant without signal pre-processing~\cite{sgv_memea2020}. In the initial evaluation, ArVE provides average errors of \unit[$1.5$]{mm} and \unit[$6.0$]{mm} in the fixed and changing $\bm{r}$ cases, respectively.

\section{Proposed Algorithm}
\label{sec:proposed_system}

The main goal of our proposal is to obtain the location of the COR as an adaptive IMU-joint vector, $\bm{r}=[r_x,r_y,r_z]^\top$, defined from the accelerometer to this COR in the sensor frame. We aim at estimating $\bm{r}$ with one IMU by using the measures of turn rate  $\bm{\omega}_I$ and specific force $\bm{f}_{A,I}$, that is the linear acceleration  $\bm{a}_{A}$ influenced by the gravity acceleration $\bm{g}$. Subindex $I$ indicates the measurements obtained directly from the IMU in its reference system.  
We obtain the IMU-joint vector $\bm{r}$ on the basis of the equation of accelerations of a rigid-solid body moving in the $3$D space~\eqref{eq:a0_Original}.
\begin{equation}\label{eq:a0_Original}
\bm{a}_0^k = \bm{a}_{A}^k + \bm{\dot{\omega}}^k \times \bm{r}^k + \bm{\omega}_I^k\times\left(\bm{\omega}_I^k\times\bm{r}^k\right),
\end{equation}
Where $\bm{a}_0^k$ and $\bm{a}_{A}^k$ are the linear accelerations in the COR and the IMU, respectively, $\bm{\omega}_I^k$ is the turn rate of the rigid-solid body and $\dot{\bm{\omega}}^k$ is its first-order derivative. As the aim is to estimate the location of fixed CORs, we assume $\bm{a}^k_0$ negligible. All parameters are expressed in the sensor frame. Superscript  $k$ denotes the time instant of parameters.
Rigid-solid bodies present a constant $\bm{r}$ vector, but in this study we focus on human bodies in which STA modify $\bm{r}^k$ at each time $k$. Besides, using~\eqref{eq:a0_Original} to estimate an adaptive $\bm{r}^k$, we assume negligible the linear acceleration caused by STA.

\begin{figure}[ht]
	\centering
	\includegraphics[width=0.79\columnwidth]{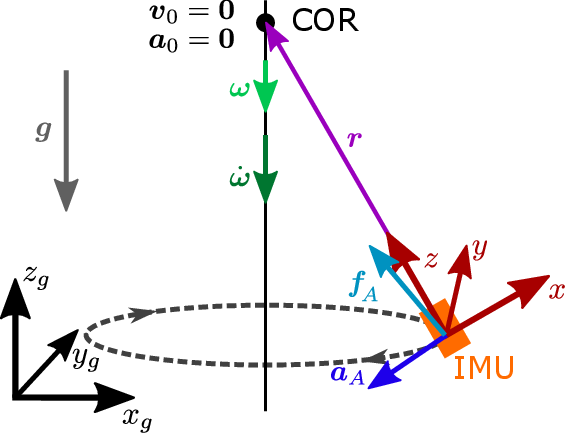}
	\caption{Scheme of the relationship between the magnitudes in~\eqref{eq:a0_Original}. The rigid-solid body moves with turn rate $\bm{\omega}_{I}$ and angular acceleration $\dot{\bm{\omega}}$, whereas the IMU suffers a linear acceleration $\bm{a_{A}}$, but it measures the specific force $\bm{f}_{A,I}$. The specific force $\bm{f}_{A,I}$ is the result of $\bm{a}_{A}-\bm{g}$ both expressed in the sensor frame. Conversely, the  linear velocity of CORs is $\bm{v}_0=\bm{0}$ by definition and as it is fixed, its linear acceleration $\bm{a}_0$ is also equal to zero.}
	\label{fig:expl_eqacels}
\end{figure}

Fig.~\ref{fig:expl_eqacels} depicts the relation of these magnitudes measured  with one IMU and the estimation of the COR. This figure shows the global frame with the subscript $g$ and the sensor frame, which is attached to the IMU.

To obtain $\bm{r}^k$ with~\eqref{eq:a0_Original}, the linear acceleration $\bm{a}_A^k$ is required. As IMUs provide the specific force $\bm{f}_{A,I}^k$ undergone by the accelerometer, we obtain $\bm{a}_A^k$ correcting the effect of the gravity through the projection of the gravity vector $\bm{g}$ into the frame of IMUs as follows:
\begin{equation}\label{eq:ftoa}
\bm{a}_A^k = \bm{f}_{A,I}^k+(C^k)^\top\bm{g},
\end{equation}
where $C^k$ is the \emph{Direction Cosine Matrix} that relates the global frame with the sensor frame and $\bm{g}$ is the gravity vector defined downwards in the global frame with a value of \unit[$9.8$]{m/s$^2$}. 
We do not use the direct measures of orientation with respect to the global frame in order to provide an algorithm usable with any generic IMU. We calculate the transformation matrix $C$ fusing the measures of turn rate  $\bm{\omega}_I^k$ and specific force $\bm{f}_{A,I}^k$ of the IMU using the algorithm introduced in~\cite{diaz2019review}. This algorithm estimates through an unscented Kalman filter (UKF) the Euler angles of the IMU from the measures of turn rate  $\bm{\omega}_I^k$ and updates these estimations with the  specific force $\bm{f}_{A,I}^k$ measured in those moments when its norm is close to the gravity vector norm.% , \unit[$9.8$]{m/s$^2$}. Thus, this algorithm corrects drift in both angles with respect to the gravity vector, which are required to eliminate the influence of this vector from the measurement of specific force $\bm{f}_{A,I}$.

In this work, we evaluate three different ways to estimate $\bm{r}$: ArVE$_d$ estimates a dynamic $\bm{r}^k$ at each time $k$ and MrVS obtains, on the one hand, an averaged $\bm{r}$ for complete tests and, on the other hand, a dynamic $\bm{r}^{k}_n$ for a determined number of samples $n$ with an overlap of \unit[$n-1$]{samples} between consecutive estimations of $\bm{r}^{k}_n$.
These methods are explained in the following two sub-sections: ArVE$_d$ in section~\ref{sec_PP:arve} and MrVS in section~\ref{sec_PP:mrvs}.

%%%%%%%%%%%%%%%%%%%%%%%%%%%%%%%%%%%%%%%%%%

\subsection{Proposed algorithm: ArVE$_d$}
\label{sec_PP:arve}

\begin{figure*}[ht]
	\centering
	\includegraphics[width=0.94\textwidth]{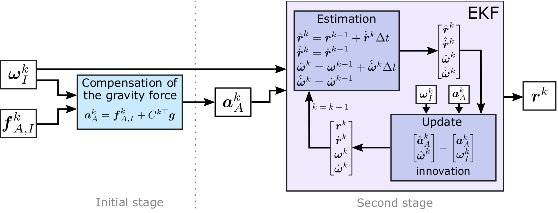}
	\caption{Flowchart to obtain the adaptive $\bm{r}^k$ at each time instant. In the initial stage, we fuse the IMU measurements of turn rate $\bm{\omega}_I^k$ and specific force $\bm{f}_{A,I}^k$ using the UKF introduced in~\cite{diaz2019review} to obtain the linear acceleration $\bm{a}_A^k$. In the second stage, we obtain $\bm{r}^k$ with this signal of linear acceleration combined with the turn rate through the EKF.}
	\label{fig:wf_frickbien}
\end{figure*}

We propose ArVE$_d$ to estimate $\bm{r}^k$ at each time instant based on the assumption of fixed CORs using an EKF. Fig.~\ref{fig:wf_frickbien} depicts the two stages of ArVE$_d$ at each time $k$: an initial stage to obtain the linear acceleration $\bm{a}_{A}^k$ followed by the second stage that consists in an EKF to determine $\bm{r}^k$. The EKF fuses the measured turn rate $\bm{\omega}_I^k$ and the calculated linear acceleration $\bm{a}_{A}^k$.

Fig.~\ref{fig:wf_frickbien} shows also the two steps of this EKF. The proposed EKF, from the second stage of ArVE$_d$, minimizes the prediction error of the state vector $\bm{x}^k$, composed of the searched oriented vector $\bm{r}^k$, its first-order derivative $\dot{\bm{r}}^k$, the turn rate $\bm{\omega}^k$ and the angular acceleration $\dot{\bm{\omega}}^k$, given the measurements from the IMU.

In the estimation step of the EKF, we assume $\hat{\dot{\bm{r}}}^k$ and $\hat{\dot{\bm{\omega}}}^k$ constant, whereas $\hat{{\bm{r}}}^k$ and $\hat{\bm{\omega}}^k$ are the integral at each time of these terms. Thus, the {estate vector} $\hat{\bm{x}}^k$ is estimated  
at each time $k$ as follows: 
\begin{equation}\label{eq:x_prediction}
\begin{cases}
\hat{\bm{r}}^k = {\bm{r}}^{k-1}+\hat{\dot{\bm{r}}}^k   \Delta t\\
\hat{\dot{\bm{r}}}^k = \dot{{\bm{r}}}^{k-1}\\
\hat{\bm{\omega}}^k = {\bm{\omega}}^{k-1} + \hat{\dot{\bm{\omega}}}^k \Delta t\\
\hat{\dot{\bm{\omega}}}^k = \dot{\bm{\omega}}^{k-1}
\end{cases}
\end{equation}

The {observations} consist of the measured turn rate  $\bm{\omega}_I^k$ and the linear acceleration $\bm{a}_{A}^k$ obtained in the initial stage of ArVE$_d$. ArVE$_d$ then updates the estimations exploiting the relationship between the estimations and the estimated linear acceleration $\hat{\bm{a}}_{A}^k$, using~\eqref{eq:updateA}, and the direct relation between the estimated turn rate $\hat{\bm{\omega}}^k$ and the measured one $\bm{\omega}^k_I$.
\begin{equation}\label{eq:updateA}
\hat{\bm{a}}_A^k = -\hat{\dot{\bm{\omega}}}^k\times\hat{\bm{r}}^k - \hat{\bm{\omega}}^k\times(\hat{\bm{\omega}}^k\times\hat{\bm{r}}^k)
\end{equation}
The linear acceleration $\bm{a}_A^k$ and the turn rate $\bm{\omega}_I^k$ are then used to obtain the {innovation} of the EKF to update the estimations at each time $k$.

Considering $\bm{\omega}^k$ and $\dot{\bm{\omega}}^k$ in the state vector, we obtain an estimation of $\dot{\bm{\omega}}^k$ using the raw data from gyroscopes, facilitating the generalized use of the algorithm. 
Since EKFs minimize the variance of the estimation error, noisy data from  IMUs do not require an initial signal filtering and avoids the post-processing suggested in~\cite{Frick2018}.  
Notice also that $\dot{\bm{r}}^k$ is not related to any measured magnitude, so it is not directly updated, but used as a parameter of adjustment of the EKF.  The use of the derivative of $\bm{r}$ in the state vector of the EKF is one of the main differences between ArVE$_d$ and ArVE, our initial approach proposed in~\cite{sgv_memea2020}, differentiated with the subindex $d$.

The  covariance parameters of the $Q$ matrix in the EKF are set according to~\cite{cerveri2003robust}. We select a constant value of covariance for each kind of tests, with synthetic and real data, indicated in section~\ref{sec_SIM:constantr_results} and section~\ref{sec_REAL:experimental_setup}, respectively, together with the explanation of the estimation of the covariance matrices $P$ and $R$.

%%%%%%%%%%%%%%%%%%%%%%%%%%%%%%%%%%%%%%%%%%
\subsection{MrVS}
\label{sec_PP:mrvs}
The original approach of MrVS proposed in~\cite{Crabolu2017} uses the complete several-second long signals of tests of the measured turn rate $\bm{\omega}_I$ and the linear acceleration $\bm{a}_A$ obtained from the measured specific force $\bm{f}_{A,I}$, and it also requires computing $\dot{\bm{\omega}}$. This parameter is obtained by discrete derivative of the turn rate $\bm{\omega}_I$ measured with the IMU. 
Since $\bm{a}_0$ is negligible in fixed CORs, the only unknown term in~\eqref{eq:a0_Original} is $\bm{r}$, so it can be rearranged as follows:
\begin{equation}\label{eq:crabolu}
\bm{a}_A = M\bm{r},
\end{equation}
where
\begin{equation}\label{eq:M}
M =  \begin{pmatrix}
-\omega_{y}^2-\omega_{z}^2 & -\dot{\omega}_z+\omega_{x}\omega_{y} &\dot{\omega}_y+\omega_{x}\omega_{z}\\
\dot{\omega}_z+\omega_{x}\omega_{y} & -\omega_{x}^2-\omega_{z}^2 & -\dot{\omega}_x+\omega_{y}\omega_{z}\\
-\dot{\omega}_y+\omega_{x}\omega_{z} & \dot{\omega}_x+\omega_{y}\omega_{z} &-\omega_{x}^2-\omega_{y}^2
\end{pmatrix}
\end{equation}
is the matrix introduced in~\cite{Crabolu2016}. Variables $\omega_x$, $\omega_y$ and $\omega_z$ are the components of the measured turn rate $\bm{\omega}_I$. In both~\eqref{eq:crabolu} and~\eqref{eq:M}, the vector $\bm{a}_A$ and the matrix $M$ symbolize a set of temporal measurements of the corresponding parameters, so no superscript $k$ is used. An averaged $\bm{r}$ is obtained solving~\eqref{eq:M} with least squares for complete tests.

When we work with several-second long IMU signals to obtain an average $\bm{r}$, the $M$ matrix from MrVS has full rank on the scenario of CORs of ball joints, as hips. However, when we look for an adaptive calculation of $\bm{r}$, uncertainties appear in~\eqref{eq:crabolu} when $\bm{\omega}$ is negligible. In these points, $M$ becomes antisymmetric, so its determinant is zero and the system is undetermined. 
Therefore, MrVS cannot be implemented in real-time applications in a straightforward way to obtain one vector per sample. Thus, we test two approaches of MrVS: obtaining an averaged $\bm{r}$ for the complete test as proposed in~\cite{Crabolu2017} and estimating an adaptive $\bm{r}^k_n$ in a sliding window with an $n$ number of samples.

%%%%%%%%%%%%%%%%%%%%%%%%%%%%%%%%%%%%%%%%%%
\section{Experiments on synthetic data}
\label{sec:sim_experiments}

We carry  out two experiments with synthetic data to test the performance of ArVE$_d$ and MrVS. The experiments simulate the motion of a pendulum moving in circles from a fixed ball joint. This pendulum imitates a limb carrying out circles from a fixed COR, as a leg moving from the hip. 
The first experiment consists in an IMU moving around a fixed COR with a constant $\bm{r}$ vector to assess the accuracy of the evaluated systems in the ideal case. 
The second experiment imitates the motion of an IMU around a fixed COR with variations of $\bm{r}$ over the test caused by simulated STA that involve small translations of the IMU. In this experiment, we study 
the error caused by assuming a constant $\bm{r}$ whereas it varies over time.

The experiments with synthetic data are presented in four sub-sections. We describe the spherical pendulum simulated to obtain the synthetic data in section~\ref{sec_SIM:spherical_pendulum} and detail the metrics used to evaluate the inertial-based methods in section~\ref{sec_SIM:metrics_errors}. Then, section~\ref{sec_SIM:constantr_results} and section~\ref{sec_SIM:variabler_results} introduce the results for these experiments carried out on synthetic data.

\subsection{Simulation of a spherical pendulum}
\label{sec_SIM:spherical_pendulum}

\begin{figure}[ht]
	\centering
	\includegraphics[width=.79\columnwidth]{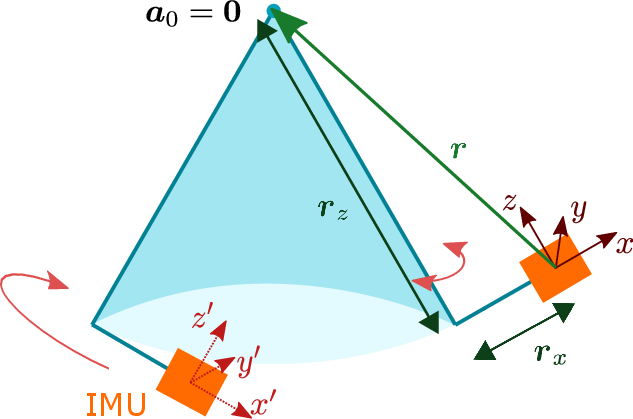}
	\caption{Scheme of the pendulum designed for simulations. The IMU (orange box) moves around the COR and its coordinate system moves with the device from positions of the initial $x$, $y$ and $z$ to each corresponding $x'$, $y'$ and $z'$. In the first experiment, $\bm{r}$ remains constant and, in the second one, its coordinates change over time.}
	\label{fig:simulations}
\end{figure}

We simulate the movement of a spherical pendulum rotating in the $3$D space during \unit[$10$]{s}, around a fixed COR and around the main axis of the pendulum. 
The pendulum describes an ellipse with two main rotations around the $x$ and $y$ axes of the simulated IMU, and a partial rotation around its $z$ axis, combining the three motions around the three IMU axes. 
The amplitudes of the movements around the $x$, $y$ and $z$ axes are \unit[$17$]{\degree}, \unit[$9$]{\degree} and \unit[$3$]{\degree}, respectively, and the motion of the pendulum around the $x$ and $y$ axes lasts \unit[$1$]{s}; and \unit[$1.5$]{s} around the $z$ axis. 
The parameters of the simulated motions are set according to the motions observed during the lower-limb calibration observed in~\cite{sgv_ipin2019}, as we do in the simulations reported in~\cite{sgv_memea2020}.
Fig.~\ref{fig:simulations} depicts these axes of the IMU together with a scheme of its motion.

We establish $\bm{r}$ considering the most likely configuration on the real scenario, where the IMU  is placed over the thigh and not in contact with the femur. In the simulation, a displacement between the main axis of the pendulum and the origin of coordinates of IMUs is taken into account and the IMU axes are misaligned with $\bm{r}$, as shown in Fig.~\ref{fig:simulations}. The $r_x$, $r_y$ and $r_z$ components are \unit[$-60$, $20$, and $200$]{mm}, respectively, so the norm of the vector is \unit[$209.8$]{mm}. 

The inertial data are simulated at a sampling rate of \unit[$100$]{Hz}. In both experiments, we add a Gaussian noise in the simulated turn rate $\bm{\omega}_I$ and specific force $\bm{f}_{A,I}$ according to the specifications of the \emph{MTw Awinda} sensors from \emph{Xsens}~\cite{xsens}, since we use these sensors in the real data experiments. 
The standard deviation of noise in the measurements of gyroscope of turn rate $\bm{\omega}$ is \unit[$0.0017$]{\degree/s}, and in the specific force $\bm{f}_A$ from the accelerometer is \unit[$0.02$]{m/s$^2$}. Bias is not considered since simulations and tests are short enough in time to be affected by it, as done in~\cite{Frick2018} and because the estimation of $\bm{r}$ does not include integration, so its estimations are insensitive to bias, according to~\cite{Seel2012}. 
In order to provide more significant results than in our previous work~\cite{sgv_memea2020}, we carry out \unit[$100$]{tests} for each experiment.

On both scenarios we set the observation noise, $R$, equally since it depends on the noise of the simulated sensors, but we adjust the estimate covariance, $P$, and the process covariance, $Q$, for each scenario.

\subsection{Metrics and errors}
\label{sec_SIM:metrics_errors}

We quantify the accuracy of the proposals using three different metrics:
\begin{enumerate}[leftmargin=*,labelsep=4.9mm]
	\item  The Euclidean norm of the vector difference between the reference $\bm{r}_r$ vector, the \emph{ground truth}, and the estimated $\bm{r}$ using the measurements from IMUs, noted with $|\Delta\bm{r}|$.  In order to consider one method competitive for its use in orientation tracking, we define the upper limit of $|\Delta\bm{r}|$ in the \unit[$10$]{\%} of the Euclidean norm of $\bm{r}_r$, because in~\cite{bonnet2015} it is reported that errors over this \unit[$10$]{\%} double errors in estimations of the orientation of limbs.
	\item The difference between the norms of  $\bm{r}_r$ and $\bm{r}$, defined as $\Delta |r|$.
	\item  The deviation angle, $\gamma$, between $\bm{r}_r$ and $\bm{r}$. 
\end{enumerate}
We consider these three metrics because each considered error has a source related with the different parameters in~\eqref{eq:a0_Original}. The difference of norms $\Delta |r|$ is mainly caused by errors in the determination of $\dot{\bm{\omega}}$.  
The deviation angle $\gamma$ is mostly affected by the accuracy of the measured linear acceleration $\bm{a}_{A,I}$ and turn rate $\bm{\omega}_I$. Finally, $|\Delta\bm{r}|$ is affected by both the difference between norms and the deviation angle.

\subsection{Results on a constant IMU-joint vector}
\label{sec_SIM:constantr_results}

Using the experiments of a simulated $3$D pendulum with a constant IMU-joint vector detailed in section~\ref{sec_SIM:spherical_pendulum}, we evaluate the accuracy of  MrVS and ArVE$_d$ to obtain an $\bm{r}$ per window and per sample, respectively.

\begin{figure}[ht]
	\centering
	\includegraphics[width=0.99\columnwidth]{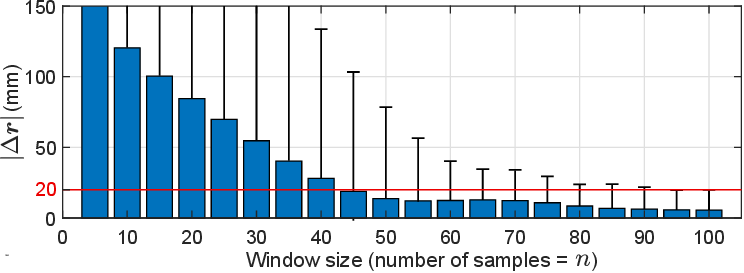}
	\caption{Average and maximum $|\Delta\bm{r}|$ of the \unit[$100$]{tests} carried out with our proposal of MrVS in a sliding window with the corresponding window size used to estimate $\bm{r}=[200, 20, -60]$\unit[$^\top$]{mm}. The horizontal red line depicts the upper limit.}
	\label{fig:mrvs_windowssize}
\end{figure}

We assess the proposal of MrVS in a sliding window as an alternative to estimate a variable $\bm{r}$ vector. We test different window sizes in order to study  the accuracy obtained with each considered number of samples $n$. 
The evaluated window sizes are from $n=5$ until \unit[$n=100$]{samples}, increasing \unit[$5$]{samples} between tests. We stop at \unit[$100$]{samples} since it would average the STA of a complete cycle in the simulations.  Windows slide \unit[$1$]{sample} to obtain each $\bm{r}$, so they overlap \unit[$n-1$]{samples}.
Since the norm of the reference vector is \unit[$209.8$]{mm}, we define the upper limit in \unit[$20$]{mm}, which is the \unit[$10$]{\%} of the vector norm. 
The resulting average $|\Delta\bm{r}|$ of each test is depicted in Fig.~\ref{fig:mrvs_windowssize}. 

Results in Fig.~\ref{fig:mrvs_windowssize} show that the errors reduce as the number of samples increases, reaching an error smaller than \unit[$10$]{mm} from the window size of \unit[$n = 80$]{samples}. The maximum errors also decrease when increasing of $n$, obtaining bearable error values under the upper limit with \unit[$n = 90$]{samples}.
However, using \unit[$95$]{samples}, the information of almost \unit[$1$]{s} is averaged, which reduces the sensitivity to changes in $\bm{r}$. 

The use of \unit[$n = 45$]{samples} in each window is a trade-off solution between the $n$ number of samples and the averaged $|\Delta\bm{r}|$ error. In this case,  the average error is  \unit[$17.6$]{mm}, which is lower than the upper limit of \unit[$20$]{mm}. Nevertheless, maximum errors are larger than \unit[$100$]{mm}.

\begin{figure}[ht]
	\centering
	\includegraphics[width=.99\columnwidth]{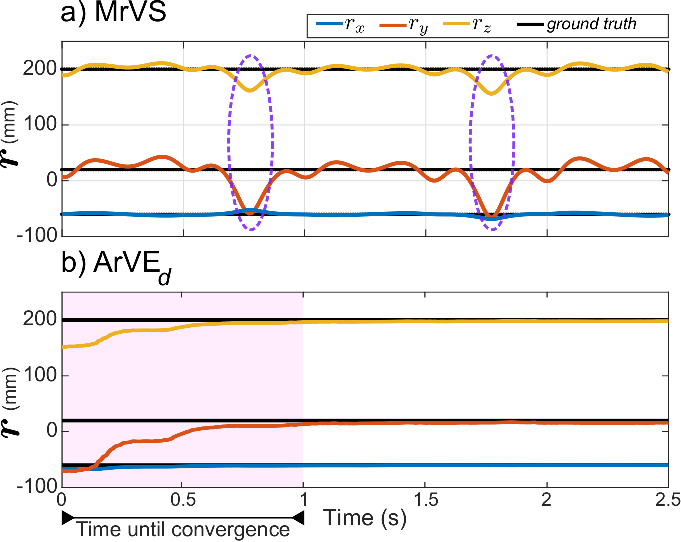}
	\caption{Results on the fixed $\bm{r}$ scenario, in which the ground truth is black depicted. a) Vector $\bm{r}^k_n$ obtained using MrVS in a sliding window of \unit[$n=45$]{samples}. b) Resulting $\bm{r}^k$ using ArVE$_d$ and setting the initial $\bm{r}_0=[0\,,0\,,0]^\top$ in the EKF. During the first second the estimations are inaccurate, until the filter convergence. After this transitory time, estimations, depicted in blue, red and yellow, are similar to the ground truth.}
	\label{fig:fixedr_2methods}
\end{figure}

Fig.~\ref{fig:fixedr_2methods}~a) 
shows the results of MrVS with a sliding window size of \unit[$45$]{samples} over the initial \unit[$2.5$]{s} of the constant $\bm{r}$ test.
The purple circles points out the intervals where errors of MrVS increase when the norm of $\bm{\omega}$ is negligible. The required number of samples to obtain an accurate estimation of the IMU-joint vector is too long to estimate a variable vector, so MrVS is not able to adapt to variations in the IMU-joint vector.

Conversely, we use ArVE$_d$ to combine the information of the IMU signals at each time instant, avoiding the inversion of the system matrix and the calculation of $\dot{\bm{\omega}}$. 
Fig.~\ref{fig:fixedr_2methods}~b) shows the resulting $\bm{r}$ vector when using ArVE$_d$ with an initial $\bm{r}_0$ composed of zeros. 
In this particular case, the EKF takes one second to converge. % when the initial $\bm{r}_0$ is set with zeros. 
After this transitory time, the estimations do not suffer from miscalculations even when $\bm{\omega}$ is close to zero. We can conclude that ArVE$_d$ provides stable estimations even in the intervals where MrVS was not able to provide an accurate result, highlighted with pink circles.

\begin{figure}[ht]
	\centering
	\includegraphics[width=0.99\columnwidth]{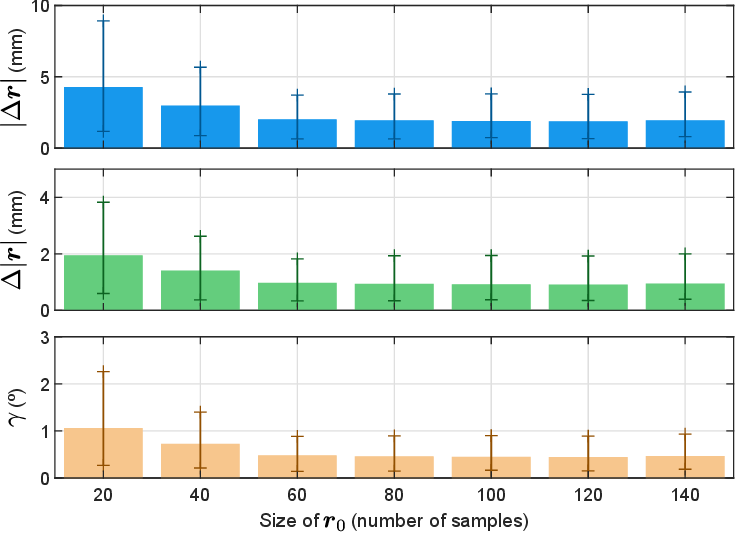}
	\caption{Errors with their corresponding ranges of values in the estimation of $\bm{r}$ with ArVE$_d$ over the \unit[$10$]{second}-experiment with the different $\bm{r}_0$ considered. The Euclidean norm of the vector difference, $|\Delta\bm{r}|$, is depicted with blue bars, the difference between norms, $\Delta|\bm{r}|$, with green bars and the deviation angle $\gamma$ in orange bars. Each bar corresponds to the average error of the \unit[$100$]{tests} carried out with this number of initial samples and the vertical lines depict the range of these errors.}
	\label{fig:setr0}
\end{figure}

Apart from the parameters of covariance in the EKF, the performance of ArVE$_d$ depends on the initial state vector, so we test the proposed method with different $\bm{r}_0$ vectors. We calculate an $\bm{r}_0$ as an average vector similarly than in MrVS, by using~\eqref{eq:crabolu} with the initial samples of tests. We use from \unit[$20$ until $140$]{samples} to estimate this $\bm{r}_0$, increasing \unit[$20$]{samples} between tests. We repeat \unit[$100$]{times} every test to evaluate the accuracy of ArVE$_d$ with each initial vector through the  metrics introduced in section~\ref{sec_SIM:metrics_errors}. Fig.~\ref{fig:setr0} depicts the average of errors, together with their maximum and minimum errors.

Fig.~\ref{fig:setr0} shows the evaluated errors and their range of values decrease as the number of samples considered to estimate $\bm{r}_0$ increases. These errors become stable when we obtain $\bm{r}_0$ with \unit[$60$]{samples}. From this number of samples, $|\Delta\bm{r}|$ is around \unit[$2$]{mm}, so using more than these \unit[$60$]{samples} (that means \unit[$0.6$]{s} of signals since \unit[$f_s = 100$]{Hz}) does not improve the accuracy of ArVE$_d$ since the EKF converges from the initial samples. For that reason, on the following we use \unit[$60$]{samples} to calculate $\bm{r}_0$ for the initialization of the EKF of ArVE$_d$. 

According to the errors shown in Fig.~\ref{fig:setr0}, ArVE$_d$ outperforms in the evaluated cases our proposal of MrVS in a sliding window. 
This improvement in accuracy is due to the fact that ArVE$_d$ has no problems with the singular points of the signal as happens with MrVS.
Despite the inaccurate estimations of MrVS near the instants when $\bm{\omega}$ is negligible, its estimations are accurate in the remaining time intervals.

\subsection{Results on variable IMU-joint vectors}
\label{sec_SIM:variabler_results}

\begin{figure*}[ht]
	\centering
	\includegraphics[width=1\textwidth]{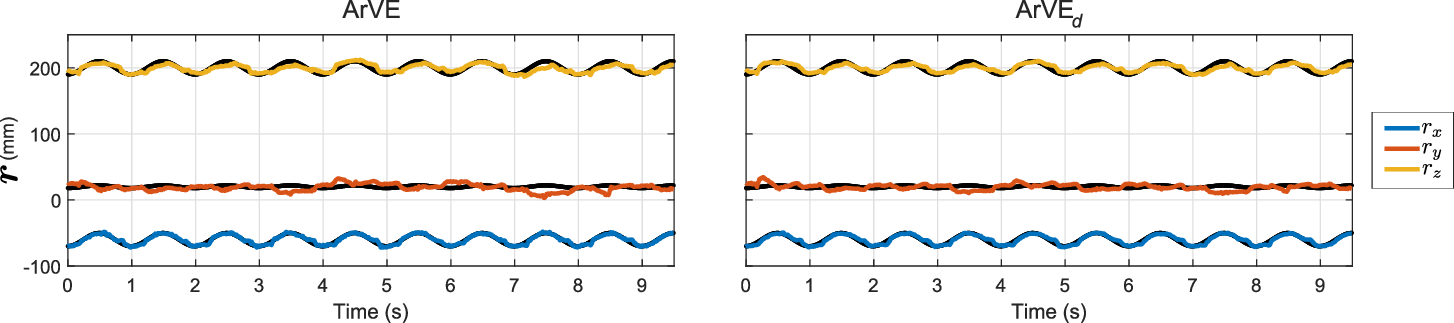}    
	\caption{Coordinates of $\bm{r}$ over the test. The ground truth is depicted in black and the estimated $r_x$, $r_y$ and $r_z$ in blue, red and yellow, respectively. We estimate $\bm{r}$ using ArVE and ArVE$_d$, and their results correspond to the images presented on the left and on the right, respectively.  }
	\label{fig:arvescomp}
\end{figure*}

We simulate the STA of the real scenario as variations in $\bm{r}$ that imitate the translation of the IMU with respect to the fixed COR. 
The variation of $\bm{r}$ over time is presented as a sinusoidal signal of frequency \unit[$1$]{Hz} and amplitude \unit[$20$]{mm} in $r_x$ and $r_z$, and \unit[$5$]{mm} in $r_y$, components previously shown in Fig.~\ref{fig:simulations}.
We set this frequency of the translational STA to make it similar to the frequency of motion of the pendulum, as suggested in~\cite{camomilla2013STA}, and the amplitude values are also set according to the results in the same work. Since IMUs are taped to the body, lateral motions over the $y$-axis are restricted, whereas the muscle contractions entail translations in the $x$-and $z$-axis.  
In this case, we compare ArVE$_d$ with ArVE to evaluate the influence of the new parameter of adjustment introduced in ArVE$_d$ on the accuracy of this proposal. 
Fig.~\ref{fig:arvescomp} depicts in black these components of the variable $\bm{r}$ vector used as ground truth over the test, together with the components of the variable $\bm{r}$ using ArVE and ArVE$_d$.

\begin{figure}
	\centering
	\includegraphics[width=1\columnwidth]{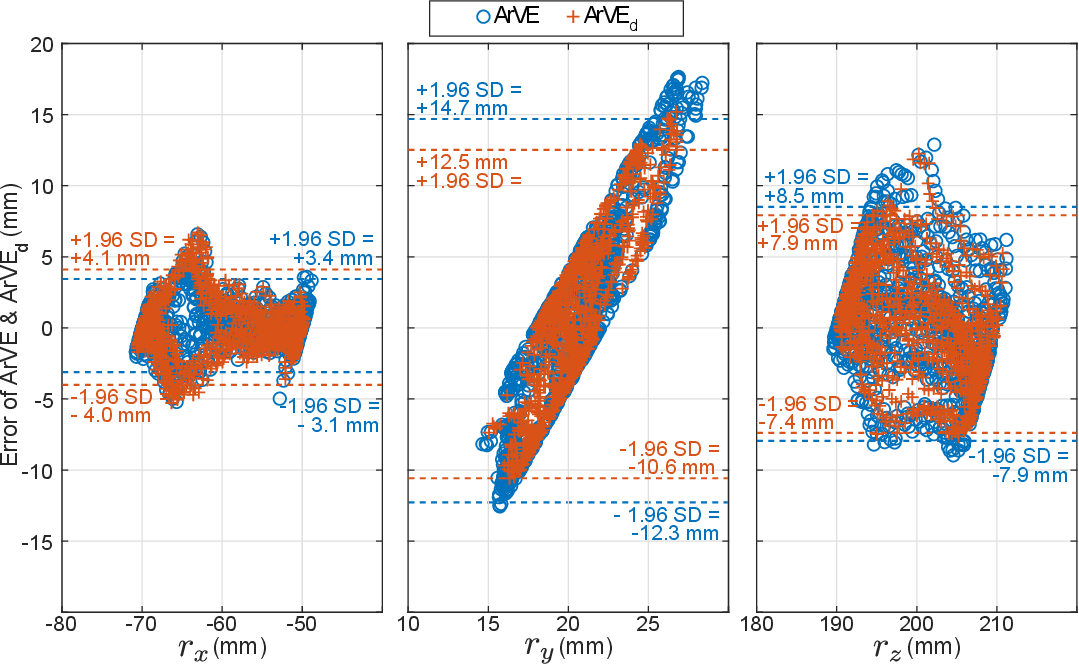}
	\caption{Bland-Altman plot with the comparison of ArVE$_d$ and ArVE. The dotted lines point out the confidence interval in which the \unit[$95$]{\%} of the errors obtained with each methods are contained, so these lines correspond with the value of  \unit[$1.96$]{times} the standard deviation (SD).}
	\label{fig:Altman_ArVEArVEd}
\end{figure}

Fig.~\ref{fig:arvescomp} shows that ArVE$_d$ and ArVE adapt to most of changes of $\bm{r}$ over time. Thus, both method provide adaptability to a variable $\bm{r}$. 
Besides, according to these results, estimations using ArVE$_d$ are closer to the ground truth, with an improvement of \unit[$7$]{\%} compared to the results obtained by using ArVE. In this way, ArVE$_d$ outperforms ArVE through the adjustment of the noise parameters of $\dot{\bm{r}}$ in the EKF.
Both methods are also evaluated by means of a Bland-Altman plot compared with the ground truth in Fig.~\ref{fig:Altman_ArVEArVEd}.

As shown in Fig.~\ref{fig:Altman_ArVEArVEd}, errors in the estimation of $r_x$ and $r_z$ are similar using ArVE$_d$ and ArVE. They are in the approximate range of \unit[$\pm4$]{mm} for $r_x$ and \unit[$\pm8$]{mm} for $r_z$. 
Conversely, there are differences in accuracy estimating $r_y$. 
The improvement in this coordinate is specially remarkable in the error dispersion, lowered \unit[$2$]{mm} in each upper-and lower-bounds. 
The results may not be as good with ${r}_y$ because its range of variation is at the noise level in the filter,  
so none of these methods adapts to its variations. 

These simulations are closer to the real scenario, where $\bm{r}$ changes due to STA, so we use the simulated data to evaluate the three methods, whose results are shown in Table~\ref{tab:errors_EKF_adaptive}. In this case, MrVS estimates a unique $\bm{r}$, constant for the complete test, whereas ArVE$_d$ and ArVE adapt to the variable vector, obtaining an instantaneous vector per sample.
\begin{table}[ht]
	\centering
	\caption{Errors in the determination of $\bm{r}$ with inertial methods with an average $\bm{r}=[200,20,-60]^\top$}
	\begin{tabular}{l|c|c|c|}
		\cline{2-4}
		& \unit[$|\Delta\bm{r}|$]{mm} & \unit[$\Delta|\bm{r}|$]{mm} & \unit[$\gamma$]{\degree} \\ \hline
		\multicolumn{1}{|p{1.2cm}|}{ArVE$_d$} & $\mathbf{6.8\pm3.9}$ & $\mathbf{3.4\pm2.4}$ & $\mathbf{1.6\pm0.9}$                     \\
		\hline
		\multicolumn{1}{|p{1.2cm}|}{ArVE}     & $7.3\pm4.6$                 & $3.8\pm2.8$                 & $1.7\pm1.1$                     \\  \hline
		\multicolumn{1}{|p{1.2cm}|}{MrVS}     & $14.4\pm3.8$                & $5.7\pm4.9$                 & $3.5\pm0.7$                     \\ \hline
	\end{tabular}\label{tab:errors_EKF_adaptive}
\end{table}

The $|\Delta\bm{r}|$ error is the most remarkable since it shows the highest improvement, from \unit[$14.4$]{mm} using MrVS, until  \unit[$6.8$]{mm} using ArVE$_d$, lowering errors more than a \unit[$50$]{\%}. 
This error decreases more than the difference of norms $\Delta|\bm{r}|$ and the angle $\gamma$ because the distance vector is affected by $\Delta|\bm{r}|$ and $\gamma$, and both errors are smaller using ArVE$_d$.  According to these results, 
ArVE$_d$ is the method that best adapts to a variable $\bm{r}$, which justifies our proposal for improvement by introducing the derivative of $\bm{r}$ in the estate vector.

It is noticeable that the errors of using ArVE$_d$ in simulations that include the simulated effect of the translational STA are similar to those presented in~\cite{Frick2018}, but they do not consider the three components in the reported errors. So, even if we cannot compare directly our results, we can conclude that our algorithm is at least as accurate as the methods in the literature. Furthermore, ArVE$_d$ only needs the initial data during \unit[$0.6$]{s} to initialize the algorithm and we get rid of the low-pass filtering of the IMU signals and the analytic derivation of the measured turn rate $\bm{\omega}_I$, used in other works as~\cite{Crabolu2016,Crabolu2017,Frick2018,Frick2018a}.

\begin{figure*}[ht]
	\centering
	\includegraphics[width=0.75\textwidth]{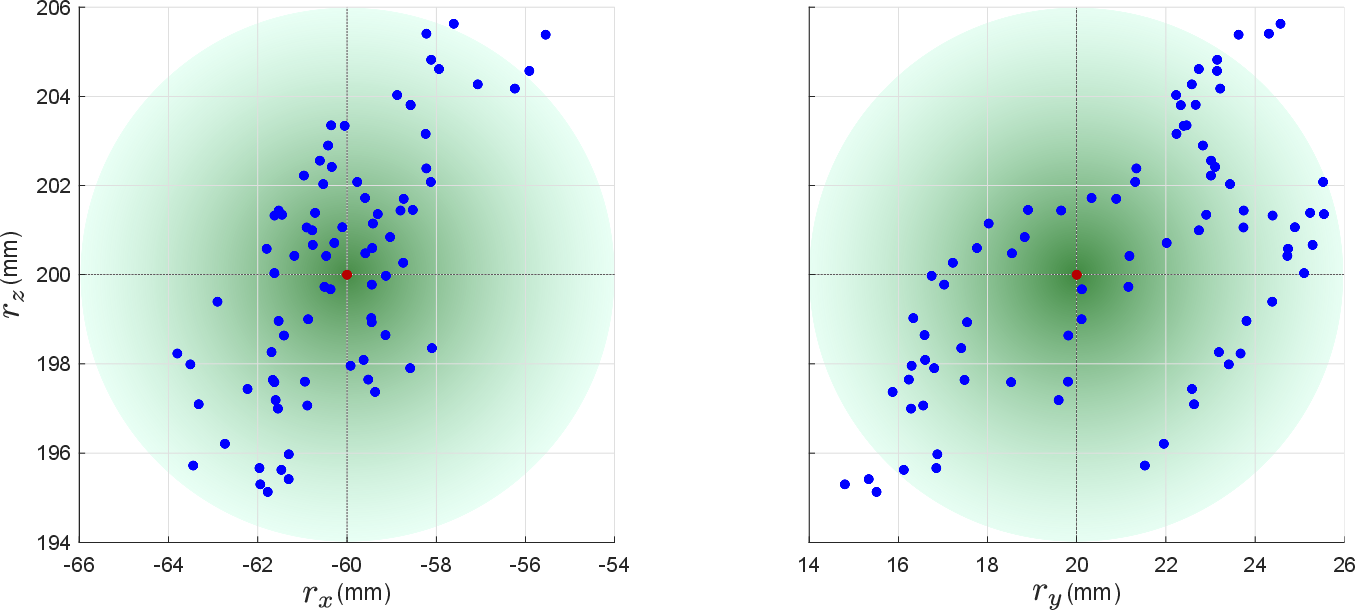}    
	\caption{Projection of the points estimated by ArVE$_d$ in planes XZ, depicted in the image on the left, and YZ, in the image on the right, over one second of the experiment. The estimated points are depicted in blue and the ground truth in red.}
	\label{fig:3D_variabler}
\end{figure*}

Finally, Fig.~\ref{fig:3D_variabler} depicts the estimated COR in the simulations of the variable $\bm{r}$ using ArVE$_d$ and, drawn in red, its actual position in two different planes. 
According to these results, the relative errors depend on the component of $\bm{r}$, being larger for the $y$-axis and the smallest variation on the $x$-axis. But it is remarkable that the \unit[$93$]{\%} of the COR estimated by ArVE$_d$ are in a sphere with a radius of \unit[$6$]{mm}. This estimated radius only is around a fifth of the hip joint radius, commonly included in the range of \unit[$25$]{mm} \unit[to $30$]{mm} according to~\cite{HipCalin2016}.

%%%%%%%%%%%%%%%%%%%%%%%%%%%%%%%%%%%%%%%%%%

\section{Experiments on the real scenario}
\label{sec:real_experiments}

We study now the performance of ArVE$_d$ on a real scenario. We obtain the COR of the hip of five volunteers with respect to one IMU, using the inertial-based systems and an optical system at the same time.
We compare ArVE$_d$ and our implementation of~\cite{Crabolu2016}, MrVS, with the optical method introduced in~\cite{DeRosario2014}.
Although other methods exist, as explained in section~\ref{sec:intro}, we use this one because it aims at obtaining a variable $\bm{r}$.

\subsection{Experimental setup}
\label{sec_REAL:experimental_setup}

\begin{figure}[ht]
	\centering
	\includegraphics[width=0.9\columnwidth]{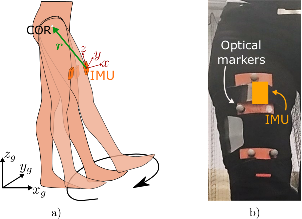}
	\caption{Experimental setup. a) Illustration of the movement performed by the volunteer in order to calibrate her/his hip. The COR, the IMU, its reference system and the global frame are shown. b) Picture of the mounting board with the IMU together with the four optical markers on the thigh of the volunteer. }
	\label{fig:leg_IMU_hip}
\end{figure}

Five volunteers with a height of \unit[$165\pm8$]{cm} participate in this study. During the experiments, they repeat \unit[$10$]{times} the hip circles motion depicted in Fig.~\ref{fig:leg_IMU_hip}~a), being equipped with one IMU placed on the thigh and four optical markers located around the IMU, as shown in Fig.~\ref{fig:leg_IMU_hip}~b). 
We choose this motion to ensure the presence of a unique COR at each time instant instead of an axis of rotation, so our system has a unique solution, that is the searched COR. The concerns about estimations of axes using~\eqref{eq:a0_Original} are more detailed in~\cite{Crabolu2018}.
The motion of hip circles is performed maintaining both legs straightened, one foot is placed on the floor while the other leg performs circles from the hip. To do this exercise, the stability of the volunteers is important to keep the hip still, so their backs rest on a stable surface and we ensure that their motions are according to the requirements for these experiments.
We eliminate the first and last signal segments of \unit[$1.5$]{s} long of tests to remove movements other than hip circles.

The inertial sensor is the \emph{MTw Awinda} from \emph{Xsens}~\cite{xsens} and the optical system consists in the Vicon equipment~\cite{vicon} together with the method proposed in~\cite{DeRosario2014} to estimate CORs. 
Both inertial and optical measurements are recorded at a sampling rate of \unit[$100$]{Hz}. 
We synchronize both systems with an initial motion of flex-extension of the hip and the following detection of the maximum position and null turn rate measured with the optical system and the IMU, respectively, in the signals measured during calibration movement. In the definition of the mounting board, we ensure that its axes are aligned with the axes of the IMU and its location center is placed at the IMU accelerometer. Using the spatial location and position of the mounting board in the reference system of the Vicon, we translate the estimations of our reference IMU-joint vector $\mathbf{v}$ into the IMU system. We use the method proposed in~\cite{DeRosario2014} because it is aimed at estimating an adaptive $\mathbf{v}$, although we finally obtain an average vector for the complete test to improve its accuracy and eliminate the dependence with the number of samples considered for each estimation. 

We obtain the location of the mounting board in which the IMU is placed, set at the location of the accelerometer in the device, its orientation and the position of each marker from the optical system. Since the IMU is aligned with the mounting board, we consider the data from the mounting board as the orientation and location of the IMU.

Besides, the covariance parameters are estimated as follows: we use the measurements of a static IMU to calculate the standard deviation needed to obtain $R$; we estimate $P$ and $Q$ using the reference data of one subject and adapting them to the best performance of ArVE$_d$, and we use these parameters for all the other subjects.

\subsection{Metrics and errors}
\label{sec_REAL:metrics_errors_real}

As optical methods are commonly used as baseline because they provide a high accuracy, we compare the outputs from both, ArVE$_d$ and our implementation of MrVS, with the results obtained trough the measurements from the optical system. We evaluate those methods with the same metrics that we used in the experiments on simulations to study the different sources of errors,  described in section~\ref{sec_SIM:metrics_errors}. In this case, our reference to determine the errors of ArVE$_d$ and MrVS is the \textbf{v} vector, obtained with the optical system and translated into the IMU system.

\subsection{Evaluation of the adaptive r on human joints}
\label{sec_REAL:ev_onhumanjoints}

\begin{figure}[ht]
	\centering
	\includegraphics[width=0.99\columnwidth]{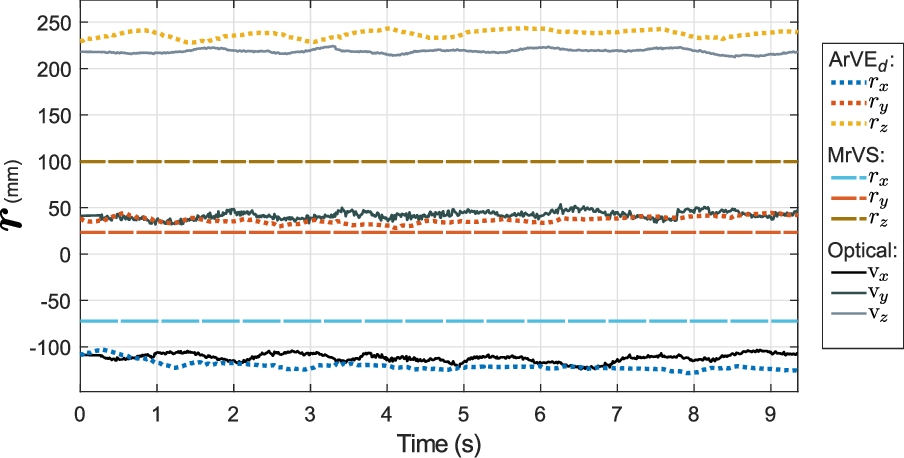}
	\caption{IMU-joint vector obtained with ArVE$_d$, depicted in blue, red and yellow dotted lines; with MrVS, presented in cyan, red and mustard stripes; and with the visual-based method, black and gray depicted in continuous lines.}
	\label{fig:cir_real_IMUvicon}
\end{figure}

On the real scenario of human hips, we have a reference \textbf{v} from the optical system, depicted in gray  in Fig.~\ref{fig:cir_real_IMUvicon}.
We use this reference to evaluate the estimations of the variable $\bm{r}$ adapted to these changes caused by the STA using ArVE$_d$ and the average $\bm{r}$ for the complete test with MrVS. 
Fig.~\ref{fig:cir_real_IMUvicon} shows that results from both adaptive algorithms, ArVE$_d$ and the optical system, experience periodic changes caused by the STA in legs during the experiment. 
This is coherent with the exercise since it consists in repetitions of the circles performed from the hip. 

\begin{figure}[ht]
	\centering
	\includegraphics[width=0.95\columnwidth]{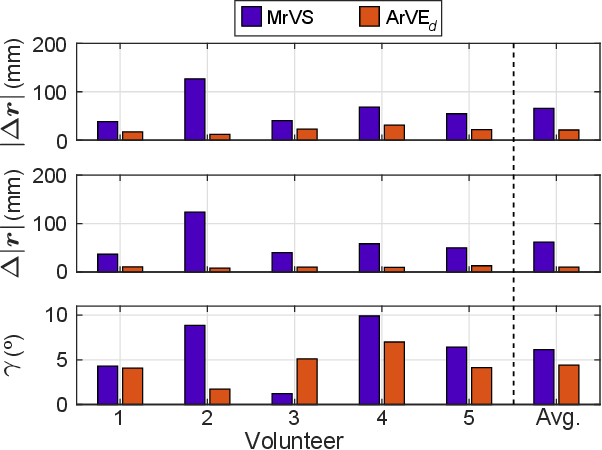}
	\caption{Errors obtained for each evaluated volunteer with MrVS and ArVE$_d$ compared with $\mathbf{v}$. The average values are also included with the label Avg.}
	\label{fig:errors_volunteers_MrVSArVE}
\end{figure}

Fig.~\ref{fig:errors_volunteers_MrVSArVE} depicts the norm of the difference vector $|\Delta\bm{r}|$, the difference of norms $\Delta|\boldsymbol{r}|$ and the deviation angle $\gamma$ of the estimations of ArVE$_d$ and MrVS with respect to $\mathbf{v}$ in the case of each evaluated volunteer. According to this figure, the errors obtained with ArVE$_d$ are similar for all volunteers and MrVS provides errors with great variability, so the accuracy of MrVS depends more on the volunteer. Also, these results show the decrement of $|\Delta\bm{r}|$ and $\Delta|\boldsymbol{r}|$ errors using ArVE$_d$ versus using MrVS in all cases and the deviation angle is similar in most cases around \unit[$5$]{\degree} with both methods, so the adaptive method is the most accurate. The differences in the average values of those errors are also consistent with this affirmation.

Table~\ref{tab:differences_real_scenario} shows the values of the errors depicted in Fig.~\ref{fig:errors_volunteers_MrVSArVE} together with their corresponding reference vector
$\mathbf{v}$ of both IMU-based approaches compared with \textbf{v} from the optical system. 
Differences exist between the metrics of both methods, being specially noticeable with respect to the difference between modulus $\Delta|\bm{r}|$ that decreases from \unit[$62$]{mm} (\unit[$28.0$]{\%}) until \unit[$10$]{mm}  (\unit[$4.5$]{\%}).
\begin{table}[{ht}]
	\centering
	\caption{Average differences between ArVE$_d$ and MrVS compared with the optical method for the five volunteers, together with the average (Avg.) values}
	\begin{tabular}{|c|c|c|c|c|c|}
		\hline Volunteer
		& $\mathbf{v}$            & Method & \unit[$|\Delta\boldsymbol{r}|$]{mm}  & \unit[$\Delta|\boldsymbol{r}|$]{mm}  & \unit[$\gamma$]{\degree} \\ \hline
		{\multirow{2}{*}{$1$}} & \multirow{2}{*}{$166\pm1$} & ArVE$_d$  & $17\pm5$  & $11\pm6$  & $4\pm2$   \\ \cline{3-6} 
		                        &                        & MrVS   & $38\pm1$  & $37\pm1$  & $4\pm1$   \\ \hline
		{\multirow{2}{*}{$2$}} & \multirow{2}{*}{$249\pm4$} & ArVE$_d$  & $12\pm2$  & $8\pm4$   & $2\pm1$   \\ \cline{3-6} 
		                       &                        & MrVS   & $126\pm3$ & $124\pm4$ &$ 9\pm1 $  \\ \hline
		{\multirow{2}{*}{$3$}} & \multirow{2}{*}{$228\pm1$} & ArVE$_d$  & $23\pm6$  & $10\pm7$  & $5\pm1$   \\ \cline{3-6} 
		                        &                        & MrVS   & $40\pm1$  & $40\pm1$  & $1\pm1$   \\ \hline
		{\multirow{2}{*}{$4$}} & \multirow{2}{*}{$235\pm2$} & ArVE$_d$  & $31\pm3$  & $9\pm6$   & $7\pm1$   \\ \cline{3-6} 
		                       &                        & MrVS   & $68\pm3$  & $58\pm2$  & $10\pm2$  \\ \hline
		{\multirow{2}{*}{$5$}} & \multirow{2}{*}{$226\pm2$} & ArVE$_d$  & $22\pm5$  & $13\pm7$  & $4\pm1$   \\ \cline{3-6} 
		                        &                        & MrVS   & $55\pm2$  & $50\pm2$  & $6\pm1$   \\ \hhline{|======|}
		{\multirow{2}{*}{Avg.}}    & \multirow{2}{*}{$221\pm2$} & ArVE$_d$  & $21\pm2$  & $10\pm3$  & $4\pm1$  \\ \cline{3-6} 
		                       &                        & MrVS   & $65\pm1$  & $62\pm1$  & $6\pm1$   \\ \hline
	\end{tabular}\label{tab:differences_real_scenario}
\end{table}

The variation between the different errors used to evaluate MrVS  is remarkable. The average deviation angle $\gamma$ versus $\mathbf{v}$ is only around \unit[$6$]{\degree}, but the error in the estimated norm $\Delta|\bm{r}|$ is \unit[$62$]{mm}, that is a \unit[$30$]{\%} of the norm of $\mathbf{v}$. This variation is a consequence of errors in the estimation of the angular acceleration $\dot{\bm{\omega}}$, since these estimations contain the propagation of errors in the measurements of turn rate $\bm{\omega}$. 
It only affects the norm since during the derivation the direction of this vector does not change.

According to Table~\ref{tab:differences_real_scenario}, we achieve a decrease in  $|\Delta\bm{r}|$ larger than a \unit[$60$]{\%} with ArVE$_d$ compared to using MrVS. 
ArVE$_d$ outperforms MrVS also in experiments on the real scenario.  
In addition, the $|\Delta\bm{r}|$ difference is in most of volunteers under the upper limit, achieving accurate results, and the averaged accuracy is  \unit[$9.5$]{\%} of the average norm of $\mathbf{v}$, which is lower than the upper limit. 

Our adaptation of MrVS obtaining an average $\bm{r}$ for the complete test entails $|\Delta\bm{r}|$ differences of \unit[$65$]{mm}, meaning a \unit[$29$]{\%} of relative error. Therefore, this method is not accurate enough to be used in the estimation of CORs with a low speed in limbs with STA. 
Differences are larger than the reported in the previous studies that use MrVS because the conditions of the evaluated tests are different, as in~\cite{Crabolu2017}. In particular, the authors evaluate the arm and the experiments are based on two perpendicular linear motions, as crosses, with a maximum turn rate of \unit[$1.6$]{rad/s}. However, our experiments are based on circular trajectories of the evaluated leg with an average turn rate of \unit[$0.8\pm0.1$]{rad/s}. This is remarkable since some of the leg exercises that may be prescribed to improve hip mobility include circular components, while not as many consider two perpendicular linear movements in a row. The speed difference is important, as it is stated in~\cite{Crabolu2016}, because errors are larger in the experiments carried out at a lower speed.

Furthermore, ArVE$_d$ is able to be implemented in real-time since it obtains one vector $\bm{r}$ per sample and only requires of the first \unit[$0.6$]{s} of tests for the initialization. The algorithms are programmed in MATLAB R2019b, running in a personal computer (processor i$7$-$8700$ at \unit[$3.2$]{GHz}, RAM memory \unit[$16$]{GB}). In this platform, the average time for the execution of a sample with ArVE$_d$ is \unit[$0.03$]{ms}. As the sampling rate is \unit[$100$]{Hz} the available time for processing a sample is \unit[$10$]{ms}. Since the execution time of a sample, i.e. the time to obtain one vector $\bm{r}$, is far lower than the sampling period, we consider that ArVE$_d$ is suitable for real-time applications, such as monitoring of rehabilitation exercises where null acceleration points exist (as in the orientation estimation of lower limbs presented in~\cite{Lin2012} or~\cite{bonnet2015}). These results of accuracy prove the usability of ArVE$_d$ as an alternative to the optical methods, being adapted to the human lower-limb scenario when the joint is constraint to be fixed.

%%%%%%%%%%%%%%%%%%%%%%%%%%%%%%%%%%%%%%%%%%
\section{Conclusions}
\label{sec:conclusions}

A novel adaptive method for IMU-joint vector determination 
is proposed and validated in this work using synthetic and real data from a hip. 
The method, called ArVE$_d$, uses raw data from an IMU to determine in real-time the COR of fixed joints with respect to the IMU location at each time instant.

With the synthetic data, ArVE$_d$ achieves an accuracy higher than \unit[$1$]{\%} of length and \unit[$1$]{\degree} of deviation when the IMU-joint vector is constant and shows adaptability to variable vectors with an error around \unit[$3$]{\%} of length and \unit[$1$]{\degree} of deviation. 
This scenario of variable vector is in which the IMU undergoes from translational movements caused by STA apart from the main rotations. 
Besides, ArVE$_d$ has also been compared with our implementation of one state-of-the-art algorithm that we have called MrVS~\cite{Crabolu2016} in this work. 
In all cases, the proposed ArVE$_d$ outperforms MrVS decreasing its errors around a \unit[$50$]{\%}. 
The accuracy of ArVE$_d$ is \unit[$10$]{\%} when is tested on real volunteers performing standardized and repetitive exercises. In this case, the reference has been obtained with an optical system. Thus, ArVE$_d$ can be considered as an alternative to estimate the IMU-joint vector, obtaining a precise COR, suitable for monitoring rehabilitation therapies that imply the motion of ball joints, as shoulders or hips.

One of the limitations of the proposal is that it is adapted to fixed joints. As future work, we are working on the adaptation of ArVE$_d$ to be applied to motions in which CORs do not show a negligible linear acceleration, such as gait or running analysis. Nevertheless, ArVE$_d$ can be used in a previous calibration step to obtain an average IMU-joint vector for off-line applications. In addition, we will adapt the algorithms to be used on portable systems, e.g. smartphones, wirelessly connected to our IMU. It will allow to provide real-time feedback in rehabilitation therapies.

\section*{Acknowledgment}
This work was supported by Junta de Comunidades de Castilla La Mancha (FrailCheck SBPLY/17/180501/000392), the Spanish Ministry of Science, Innovation and Universities (MICROCEBUS RTI2018-095168-B-C51) and the Youth Employment Program (PEJ-2017-AI/TIC-7372). 
The authors would like to thank the Deutschen Zentrums fur Luft-und Raumfahrt (DLR) for its collaboration in the measurement campaign and the borrowing of the necessary infrastructure and equipment.

\section*{Abbreviations}
The following abbreviations are used in this manuscript:\\
	\begin{tabular}{@{}ll}
	ArVE & adaptive r vector estimator\\
	ArVE$_d$ & adaptive r vector estimator, considering $\dot{\bm{r}}$\\
	MrVS & mean r vector least-squares-based estimator\\
	COR & center of rotation\\
	STA & soft tissue artifacts\\
	IMU & inertial measurement unit\\
	EKF & extended Kalman filter\\
	UKF & unscented Kalman filter \\
	SD & standard deviation
	\end{tabular}

\ifCLASSOPTIONcaptionsoff
  \newpage
\fi

\bibliography{bibliography}
\bibliographystyle{IEEEtran}

\end{document}